\documentclass[aps,pre,longbibliography,notitlepage,floatfix,nofootinbib]{revtex4-1}

\usepackage{bm,amsmath,amssymb,graphicx,stmaryrd,float,subcaption,upgreek,sidecap}
\usepackage[abs]{overpic}
\newcommand{\uvc}[1]{\bm{\mathrm{\hat #1}}} %unitvector, bold with hat, not slanted
\usepackage{comment,footnote}
\usepackage[format=plain,labelfont={bf,small},textfont=small,justification=raggedright,singlelinecheck=false]{caption}

\usepackage{mathrsfs}

\usepackage[T1]{fontenc} %for {\k{e}} in one of the references

\newcommand{\bX}{{\bf X}}
\newcommand{\bP}{{\bf P}}

\newcommand{\bF}{{\bf F}}

\newcommand{\bv}{{\bf v}}

\newcommand{\bA}{{\bf A}}
\newcommand{\bB}{{\bf B}}
\newcommand{\bg}{{\bf g}}
\newcommand{\bI}{{\bf I}}
\newcommand{\norm}[1]{#1\cdot#1}
\newcommand{\jump}[1]{\llbracket#1\rrbracket}
\newcommand{\add}[1]{\{#1\}}

\begin{document}
	\title{Pick-up and impact of flexible bodies}
	\author{H. Singh}
	\email{harmeet@vt.edu}
	\affiliation{Department of Biomedical Engineering and Mechanics, Virginia Polytechnic Institute and State University, Blacksburg, VA 24061, U.S.A.}
	\author{J. A. Hanna}
	\email{hannaj@vt.edu}
	\affiliation{Department of Biomedical Engineering and Mechanics, Department of Physics, Center for Soft Matter and Biological Physics, Virginia Polytechnic Institute and State University, Blacksburg, VA 24061, U.S.A.}

	\date{\today}

\begin{abstract}

Picking up, laying down, colliding, rolling, and peeling are partial-contact interactions involving moving discontinuities.
We examine the balances of momentum and energy across a moving discontinuity in a string, with allowance for injection or dissipation by singular supplies.
We split the energy dissipation according to its invariance properties, discuss analogies with systems of particles and connections with the literature on shocks and phase transition fronts in various bodies, and derive a compatibility relation between supplies of momentum and translation-invariant energy.
For a moving contact discontinuity between a string and a smooth rigid plane in the presence of gravity, we find a surprising asymmetry between the processes of picking up and laying down, such that steady-state kinks in geometry and associated jumps in tension are not admissible during pick-up.  This prediction is consistent with experimental observations.
We briefly discuss related problems including the falling folded chain, peeling of an adhesive tape, and the ``chain fountain''.  Our approach is applicable to the study of impact and locomotion, and to systems such as moored floating structures and some musical instruments that feature vibrating string and cable elements interacting with a surface.

\end{abstract}

\maketitle

\section{Introduction}

Picking up or laying down a floppy object is not a particularly exotic activity, and one might assume the process to be well understood.  However, dynamic interactions between flexible and rigid bodies are often counterintuitive.
A chain falling onto a table experiences an additional acceleration beyond that of free-fall \cite{HammGeminard10,Grewal11}, ropes and chains lift off of pulleys or other guiding surfaces \cite{Rennie72, Calkin89, Cambou12, Brun16}, and form arches \cite{HannaKing11, MouldSiphon2, Biggins14} when pulled from piles at rest.
In perfectly flexible objects such as idealized strings and some real chains, shape discontinuities--- kinks--- can form.  Contact with obstacles or guides can help sustain such jumps in the geometry, and in other fields such as the axial tension, momentum, and energy.

In the present work, we consider a propagating discontinuity separating contacting and free regions of an inextensible string.
We examine a commonplace and seemingly simple process, that of picking up or laying down a flexible extended body from or onto a rigid surface.  
We find a surprising asymmetry in this problem, namely that kinks and associated jumps in tension are admissible in lay-down impact onto, but generally inadmissible in pick-up from, a smooth surface.
In lifting a string or chain from a table or the floor, one observes that its tangents are continuous.  That this is a necessity can be determined from the jump conditions for linear momentum and energy %(per time per area)
across the contact discontinuity.  No such requirement exists for impact.  
This prediction is consistent with our own limited experimental evidence, as well as experiments and theoretical results from other workers on linearized and related problems.
The dynamics we describe arise in deployment, unspooling, and lay-down of space tethers, yarn, machine belts, and submarine cables \cite{BeletskyLevin93, BatraFraser15, IrschikBelyaev14, Zajac57}, as well as vibration of structures as disparate as the strings of the \emph{veena} \cite{Raman21, Burridge1982}, catenary moorings of offshore structures \cite{Gobat2001}, and peeling adhesive tapes \cite{Rivlin44, Kendall75, BurridgeKeller1978, Ericksen1998, Pede_peeling2006, Cortet13, Dalbe14}.

% liquid rope coiling Ribe12 ?

%Beatty and Haddow\cite{BeattyHaddow1985} 

Our approach follows, and modifies, previous work on the mechanics of discontinuities in strings and rods.  A general treatment of shocks in one dimensional media, starting from Green and Naghdi's framework \cite{GreenNaghdi78}, was undertaken by O'Reilly and Varadi \cite{OReillyVaradi99,OReillyVaradi03}, further developed by O'Reilly \cite{OReilly2007,OReilly07}, and applied by Virga \cite{Virga14,Virga15} to several counterintuitive problems in chain dynamics.
 One important aspect of this approach is the inclusion of singular source terms in jump conditions for field quantities.  Such terms allow for the injection of momentum and dissipation of energy by interaction with an obstacle, or other means.  It is the relationships between these terms that will lead us to our results.
A thermo-mechanical treatment of moving discontinuities in elastic bars, which are kink-free by definition, may be found in Ericksen \cite{Ericksen1998}.  The work of Abeyaratne and Knowles  \cite{Abeyaratne1990, Abeyaratne1991} on propagating phase transition fronts in bulk solids and bars was extended to strings, which do admit kinks, by Purohit and Bhattacharya \cite{Purohit2003}.  Source terms do not appear in these treatments.
Podio-Guidugli and co-workers \cite{Pede_peeling2006} treat the problem of peeling by employing a configurational approach, including a source-like term.
%but always alongside their conjugate quantities, which seems to avoid some of the ambiguity that leads to differences in the present treatment and those of O'Reilly\cite{OReilly2007,OReilly07} and Virga\cite{Virga14,Virga15}.

We begin with a derivation of the governing equations and the relevant jump conditions in Section \ref{governing}.  We also construct a natural orthogonal co-ordinate system at the discontinuity that proves useful in subsequent manipulations.
In Section \ref{invariance}, we isolate the translation-invariant power dissipated at the discontinuity, and interpret this through two analogies with particle impact systems. %, and show that our definition is equivalent to measuring the energy of a system from its center of mass.  
 The functional form of this invariant power is shown to be consistent with the notion of a driving traction proposed by Abeyaratne and Knowles \cite{Abeyaratne1990,Abeyaratne1991}.
We also contrast our definition with those proposed by O'Reilly \cite{OReilly2007,OReilly07} and Virga \cite{Virga14,Virga15}. 
In Section \ref{compatibility}, we introduce a compatibility relation between linear momentum and energy source terms.  %Among other things, this relation reveals the inconsistency of some assumptions made in the literature on space-fixed loads on axially moving structures.  
In Section \ref{conserving}, we discuss conditions for the conservation of translation-invariant energy.
In Section \ref{admissibility}, we use our compatibility relation to determine the necessary conditions for a kink to exist in a particular physical situation of interest.
We consider pick-up and impact involving an inextensible, perfectly flexible one-dimensional body and a smooth, flat, frictionless surface.   
We require that the jump in translation-invariant energy at the contact discontinuity be non-positive, and conclude that a steady-state kink may exist at the point of impact but not at pick-up.  Some experimental evidence in support of this prediction is presented involving an axially moving chain perturbed into a ``bubble'' on a surface, as well as prior work by Gobat and Grosenbaugh \cite{Gobat2001} on the touchdown point of vibrated pendant chains.
The appendices provide an alternative justification for the non-positivity of the power injection in pick-up and impact that employs a form of the Clausius-Duhem inequality, as well as brief discussions of related problems: a falling folded chain, peeling of an adhesive tape, and deployment from a pile, including the ``chain fountain''.

\section{Geometry and governing equations}\label{governing}

Consider a one-dimensional body described as a curve $\bX(s,t)$ parameterized by its arc length $s$ and the time $t$.  We will be concerned with inextensible strings, the simplest example of such a body.  The assumption of inextensibility means that the arc length $s$ is also a material coordinate.  Let there be a discontinuity in any or all field quantities at the point $s=s_0(t)$.
The sum and difference (jump) of a field quantity across the discontinuity are denoted by  $\add{Q}\equiv Q^+ +Q^-$ and $\jump{Q}\equiv Q^+-Q^-$, respectively,\footnote{Our notation differs by a factor of two from that of O'Reilly \cite{OReillyVaradi99,OReillyVaradi03,OReilly2007,OReilly07}, who uses $\add$ to denote the average.} where $Q^\pm$ are the values of $Q$ on either side of the point $s_0$.  
Nearly all of the equations in this paper will be jump conditions holding only at the non-material point $s_0$, which point we will refer to as a discontinuity, jump, or shock.

The following kinematic constraints on position and velocity hold at the shock,
\begin{align}
\jump{\bX}&=0\, ,\label{position_compatibility}\\
\jump{\partial_t\bX+\partial_t s_0 \partial_s\bX}&=0\, ,\label{velocity_compatibility}
\end{align}
where \eqref{position_compatibility} is simply a statement that the curve is continuous, and the velocity compatibility condition \eqref{velocity_compatibility} ensures continuity of the spatial velocity at the non-material point of discontinuity.
Condition \eqref{velocity_compatibility} is obtained by taking a total time derivative of $\jump{\bX(s_0(t),t)}$.  In our notation, $\partial_t$ indicates a material time derivative (constant $s$) when applied to the position vector $\bX(s,t)$, and a partial time derivative when applied to $s_0(t)$.\footnote{Where we write $\partial_t s_0$, Virga \cite{Virga14,Virga15} writes $\dot{s}_0$ and O'Reilly \cite{OReillyVaradi99,OReillyVaradi03,OReilly2007,OReilly07} writes $\dot{\gamma}$.}  Condition \eqref{velocity_compatibility} links spatial velocities of material points $\partial_t\bX^\pm$, tangent vectors $\partial_s\bX^\pm$, and the material velocity of the shock $\partial_t s_0$.
Due to the continuity of the coordinate $s$, $\partial_t s_0$ commutes with the double bracket, and so can be moved in or out of it. 
 
\begin{figure}[h!]
	\includegraphics[width=8cm]{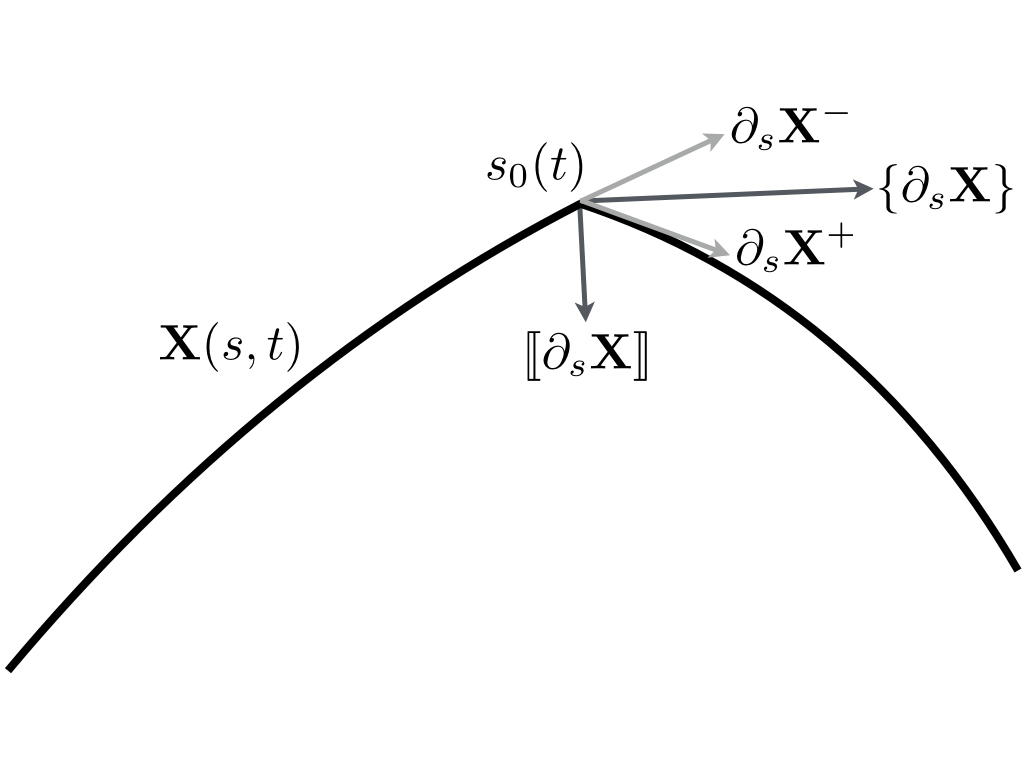}
	\captionsetup{margin=3cm}
	\caption{A one-dimensional body $\bX(s,t)$ with a discontinuity at $s=s_0(t)$.  Also shown are unit tangent vectors $\partial_s\bX^\pm$ on either side of the discontinuity, and their sum $\add{\partial_s\bX}$ and jump $\jump{\partial_s\bX}$.}
	\label{string_with_discontinuity}
\end{figure}

That $s$ is also an arc length coordinate means that $\partial_s\bX^\pm$ are unit vectors, and thus their sum and jump form an orthogonal pair: $\jump{\partial_s\bX}\cdot\add{\partial_s\bX}=0$.
As long as the tangents are neither continuous ($\jump{\partial_s\bX}=0$) nor perfectly folded-back ($\add{\partial_s\bX}=0$), this pair of axes is quite useful at the point of interest, and we will often break vector quantities into their projections onto the sum and jump of tangents.  Figure \ref{string_with_discontinuity} shows the geometry of the relevant objects.
Projections of the velocity compatibility condition \eqref{velocity_compatibility} reveal relationships between the geometry and the material and spatial velocities,
\begin{align}
	&\jump{\partial_t\bX}\cdot\add{\partial_s\bX} =0\, ,\label{velocity_constraint} \\
	&\jump{\partial_t\bX}\cdot\jump{\partial_s\bX} = -\partial_t s_0\,\norm{\jump{\partial_s\bX}} \, . \label{shock_speed}
\end{align}
We will often make use of the identity
\begin{align}
2\jump{\bA\cdot\bB}=\jump{\bA}\cdot\add{\bB}+\add{\bA}\cdot\jump{\bB} \label{identity}
\end{align}
for two vectors $\bA$ and $\bB$.

Our discussion will center on two additional jump conditions that may be derived in different ways, one way being the approach of O'Reilly and Varadi \cite{OReillyVaradi99}.  We briefly outline the different approach of a prior paper \cite{Hanna15}, which employs actions with time-dependent boundaries of the form
\begin{align}
	A &= \int \!\!dt \int^{s_0(t)} \!\!ds \,\mathcal{L} \, , \label{action}\\
	2\mathcal{L} &= \mu\partial_t\bX\cdot\partial_t\bX - \sigma\left(\partial_s\bX\cdot\partial_s\bX - 1\right) \, , \label{lagrangian_density}
\end{align}
one such action for each side of the jump.  The quantity $\mathcal{L}$ is the constrained kinetic energy density of an inextensible string, with the constant $\mu$ a uniform mass density and the multiplier $\sigma(s,t)$ the tension.
Variation of $\bX$ leads to the string equations in the bulk,
\begin{align}
 \mu\partial^2_t\bX-\partial_s(\sigma\partial_s\bX)=0\, ,\label{bulk_momentum_equation} 
\end{align}
while variations of $\bX$ and $t$ lead to jump conditions at the shared moving ``boundary'' $s=s_0(t)$, 
\begin{align}
\bP+\jump{\sigma\partial_s\bX+\mu\partial_t s_0\partial_t\bX}&=0\, ,\label{momentum_jump}\\
\tilde{E}+\jump{\sigma\partial_s\bX\cdot\partial_t\bX+\tfrac{1}{2}\mu\partial_t s_0\partial_t\bX\cdot\partial_t\bX}&=0\, ,\label{energy_jump}
\end{align}
where, following Green and Naghdi \cite{GreenNaghdi78} and O'Reilly and Varadi \cite{OReillyVaradi99}, we allow for singular sources of stress $\bP$ and power per area $\tilde{E}$ that represent in general form the effect of obstacles, external forces, internal dissipative processes, and so on, that are not explicitly accounted for in the actions.  We will loosely refer to these jump conditions as those for (linear) momentum and energy, though the correct units are these quantities per time per area.
A negative value of energy injection rate $\tilde{E}$ means dissipation of energy.\footnote{There are sign differences in our present definitions of singular sources and those in \cite{Hanna15}.}  This quantity is not invariant under translations of the system.

We can express the momentum jump condition \eqref{momentum_jump} in the sum-jump basis mentioned above. 
Using the velocity compatibility condition \eqref{velocity_compatibility}, we rewrite  \eqref{momentum_jump} as
\begin{align}
\bP=-\jump{(\sigma-\mu\partial_t s_0^2)\partial_s\bX}\, , \label{momentum_jump_modified} 
\end{align}
and then use the identity \eqref{identity} to expand this as
\begin{align}
\bP=-\tfrac{1}{2}\jump{\sigma}\add{\partial_s\bX}-\tfrac{1}{2}\add{\sigma-\mu\partial_t s_0^2}\jump{\partial_s\bX}\, .\label{momentum_jump_basis}
\end{align}
We will make use of this expression in subsequent discussion.

\section{Translational invariance and energy dissipation}\label{invariance}

In this section, we isolate a translation-invariant measure of the dissipation of energy at the shock. This quantity is equivalent to what was introduced in an \emph{ad hoc} manner in prior work \cite{Hanna15}.
The term $\tilde{E}$ in the energy jump condition \eqref{energy_jump} is the total areal power %(energy per time per area) 
input or loss across the shock in some choice of inertial frame.  
Because some of this is associated with changes in kinetic energy, $\tilde{E}$ is not invariant under translations.  We extract the translation-invariant part by using the identity \eqref{identity} to rearrange  \eqref{energy_jump} in terms of the (generally non-orthogonal) sum and jump of velocities, then rewrite using the momentum jump condition \eqref{momentum_jump},
\begin{align}
0 %&= \tilde{E}+\jump{\sigma\partial_s\bX\cdot\partial_t\bX}+\tfrac{1}{2}\mu\partial_t s_0 \jump{\partial_t\bX\cdot\partial_t\bX}\, ,\\
&= \tilde{E}+\tfrac{1}{2}\jump{\sigma\partial_s\bX+\mu\partial_t s_0\partial_t\bX}\cdot\add{\partial_t\bX}+\tfrac{1}{2}\add{\sigma\partial_s\bX}\cdot\jump{\partial_t\bX}\, ,\label{energy_rewritten_1}\\
&=\tilde{E}-\tfrac{1}{2}\bP\cdot\add{\partial_t\bX}+\tfrac{1}{2}\add{\sigma\partial_s\bX}\cdot\jump{\partial_t\bX}\, .\label{energy_rewritten_2}
\end{align}
The total power is split into a non-invariant piece which consists of the external stress $\bP$ acting on the average spatial velocity of material points, and an invariant piece which is the negative of the average contact force acting on the jump in this velocity.  In the following subsections, we suggest interpretations for these quantities.  We can loosely think of the non-invariant term as the external working on the material, and the invariant term as some kind of ``internal'' dissipation.  

%We note that for our inextensible, flexible string, there is only kinetic energy. 

The splitting of the power in \eqref{energy_rewritten_2} differs from that of Podio-Guidugli \cite{Podio1997inertia}, who defines the inertial power as that which changes the total kinetic energy.  This is because the kinetic energy can also be split according to its invariance properties.
While the kinetic energy of a system depends on the choice of frame, the changes in kinetic energy due to internal collisions are frame invariant (as shown, for example, in Section 96 of Whittaker \cite{Whittaker88}).

We define the translation-invariant energy injection rate as
\begin{align}
E = \tilde{E} - \tfrac{1}{2}\bP\cdot\add{\partial_t\bX}\, ,\label{energy_invariant_definition}
\end{align}
the difference between the rate of change of the total energy of the shock and the working of external forces on the infinitesimal interval of the string that encloses the shock \cite{Abeyaratne1990,Abeyaratne1991}. This form of the power injection also arises when applying a form of the Clausius-Duhem inequality to isentropic, isothermal, purely mechanical systems, as shown in Appendix \ref{clausius_duhem}.  
Note that for a rod with bending energy, a non-rotationally invariant term involving a source of areal torque and the rates of change of the tangent vectors would appear as well.

\subsection{Two examples involving discrete particles}\label{particle_examples}

To illustrate some of the relevant concepts, we present two examples involving systems of  rigid particles.

Consider first\footnote{We thank J. S. Biggins for suggesting this analogy.} the total kinetic energy of such a system, which may be written in a special way using the center of mass velocity $\bv_{com}$:
\begin{align}
\tilde{T} = \sum_{j=1}^{N}\tfrac{1}{2} m_j (\bv_j-\bv_{com})^2+\tfrac{1}{2}m_j \bv_{com}\cdot\bv_{com}\, .\label{kinetic_energy_split}
\end{align}
The first term on the right side is the kinetic energy of the particles as observed from the frame of the center of mass of the system, whereas the second term is the kinetic energy associated with the center of mass itself. The total kinetic energy $\tilde{T}$ in (\ref{kinetic_energy_split}) is not invariant under translations, but the quantity $T \equiv \tilde{T} -\big(\sum_{j=1}^{N}\tfrac{1}{2}m_j\big) \bv_{com}\cdot\bv_{com}$ is.  Taking a total time derivative of this quantity, we obtain
\begin{align}
\frac{d}{dt}(T)=\frac{d}{dt}(\tilde{T})-\bF_{ext}\cdot \bv_{com}\, ,\label{kinetic_energy_analogy_final}
\end{align}
where $\bF_{ext}$ is any external force effectively acting on the center of mass.
The invariant term on the left side is the rate of change of energy as observed from the center of mass of the system of particles. We may trivially identify terms in (\ref{kinetic_energy_analogy_final}) and (\ref{energy_invariant_definition}).  The quantity $E$ is analogous to the rate of change of energy as observed from the center of mass of the infinitesimal interval on the string that encloses the shock.  This is the energy available for ``internal'' dissipation by \emph{e.g.} particle-particle or chain link-link collisions, without the aid of external forces.
The derivation above holds equally well if the center of mass frame is accelerating. A detailed discussion on the special properties of the center of mass frame and the validity of the work-energy theorem in inertial and non-inertial frames of reference can be found in \cite{Diaz2008}.

A second example will help explain the form of the external working in \eqref{energy_rewritten_2} and \eqref{energy_invariant_definition}.  In these expressions, the external stress $\bP$ is associated with the average material velocity $\tfrac{1}{2}\add{\partial_t\bX}$ across the shock. This pairing also arises during the collision of two rigid bodies, where the working of the impact is the product of the impulse with the mean of the relative velocities describing approach and separation.  The derivation we reproduce here is essentially that of Article 308 of Thomson and Tait \cite{ThomsonTait1879}, and is an elementary result in impact mechanics \cite{Stronge2004impact}.  
The change in kinetic energy due to a two-particle collision is
\begin{align}
\Delta\tilde{T} &= \tfrac{1}{2} m_1 (\norm{\bv_{1f}}-\norm{\bv_{1i}}) + \tfrac{1}{2} m_2 (\norm{\bv_{2f}}-\norm{\bv_{2i}})\, , \label{kinetic energy change}\\
 &= \tfrac{1}{2} m_1 (\bv_{1f} - \bv_{1i})\cdot(\bv_{1f} + \bv_{1i}) + \tfrac{1}{2} m_2 (\bv_{2f} - \bv_{2i})\cdot(\bv_{2f} + \bv_{2i})\, ,\label{kinetic_energy_change_2} \\
&= \bI\cdot\tfrac{1}{2}(\bv_{app}+\bv_{sep})\, ,\label{kinetic_energy_change_final}
\end{align}
where the symbols for initial and final velocities of particle 1 and particle 2 are straightforward, $\bv_{app} \equiv \bv_{2i} - \bv_{1i}$ and $\bv_{sep} \equiv \bv_{2f} - \bv_{1f}$ are the approach and separation velocities, and $\bI = m_2(\bv_{2f} - \bv_{2i}) = - m_1(\bv_{1f} - \bv_{1i})$ is the impulse experienced by each particle due to the collision.
It is the average of the approach and separation velocities that appears conjugate to the impulse.  Our problem corresponds to the case where one ``particle'' is a stationary obstacle.

Our definition of the working of the source of momentum $\bP$ differs from the approaches taken by O'Reilly \cite{OReilly07} and Virga \cite{Virga15}, who associate $\bP$ with the spatial velocity of the non-material point $\bX(s_0)$ corresponding to the shock. We provide further detail on their choice in the following subsection.

\subsection{Rearrangements and relationships}\label{furtherstuff}

Returning to the energy jump in the form \eqref{energy_rewritten_2}, and the definition \eqref{energy_invariant_definition}, we have
\begin{align}
E = -\tfrac{1}{2}\add{\sigma\partial_s\bX}\cdot\jump{\partial_t\bX}\, .\label{energy_invariant_raw}
\end{align}
Employing velocity compatibility \eqref{velocity_compatibility}, we may write
\begin{align}
E=\partial_t s_0\tfrac{1}{2}\add{\sigma\partial_s\bX}\cdot\jump{\partial_s\bX}\, .\label{energy_measure_velocity_substituted}
\end{align}
Essentially this form for the dissipation at a moving shock can be found in the work of Abeyaratne and Knowles \cite{Abeyaratne1990, Abeyaratne1991}. They proposed the concept of a ``driving traction'' as a dynamic conjugate to the material velocity $\partial_t s_0$ of a strain discontinuity in a bulk solid or unkinked, extensible bar.  If the driving traction vanishes everywhere on the surface of discontinuity, the motion is dissipation-free.  Purohit and Bhattacharya \cite{Purohit2003} derived a thermomechanical expression for the driving force associated with a moving phase boundary, \emph{sans} source terms, in a kinked string made of a phase-transforming material. This coincides with our expression \eqref{energy_measure_velocity_substituted} in the inextensible, purely mechanical case.\footnote{When the source $\bP = 0$, \eqref{identity} and \eqref{momentum_jump_modified} can be used to simplify the driving traction from $\tfrac{1}{2}\add{\sigma\partial_s\bX}\cdot\jump{\partial_s\bX}$ to just $\jump{\sigma}$.}

Using \eqref{identity}, the expression \eqref{energy_measure_velocity_substituted} can be rewritten in an informative way in terms of the jumps in tension and tangents,
\begin{align}
E=\tfrac{1}{4}\partial_t s_0 \jump{\sigma}\norm{\jump{\partial_s\bX}}\, .\label{energy_measure_final}
\end{align}
This expression indicates that a propagating shock conserves translation-invariant energy if either the tensions or the tangents are continuous across the shock.  A similar observation about the tensions was made by Calkin and March \cite{CalkinMarch89} for the case of a falling folded chain, where \eqref{energy_measure_final} reduces to $E=\partial_t s_0 \jump{\sigma}$, in agreement with equation (12) of \cite{CalkinMarch89}.\footnote{In that problem, $\norm{\jump{\partial_s\bX}}=4$, and the $\dot{x}$ in equation (12) of \cite{CalkinMarch89} is twice the shock speed, $\dot{x}=2\partial_t s_0$.}  % due to the velocity compatibility condition (\ref{velocity_compatibility}).}.  
We discuss that problem in Appendix \ref{falling}.

The energy supply at the shock defined by \eqref{energy_measure_final} differs from equation (2.25) of Virga \cite{Virga15} by the geometric factor $\tfrac{1}{4}\norm{\jump{\partial_s\bX}}$. This difference arises due to Virga's choice of the spatial velocity $\bv_0$ of the non-material point of discontinuity,
\begin{align}
\bv_0=\partial_t\bX^\pm+\partial_t s_0 \partial_s\bX^\pm = \tfrac{1}{2}\add{\partial_t\bX} + \tfrac{1}{2} \partial_t s_0 \add{\partial_s\bX}\, , \label{spatial_velocity}
\end{align}
as conjugate to the external force $\bP$, in contrast to our choice of the average spatial velocity of material points $\tfrac{1}{2}\add{\partial_t\bX}$.  From the above expression, we note that this average is simply the portion of  $\bv_0$ that is not translation-invariant.
Although the energy supply derived in \cite{Virga15} does not display any dependence on the the jump in tangents, Virga assumes a constitutive law in equation (2.27) of \cite{Virga15} which does.  In our derivation, the geometric dependence arises naturally from our splitting of the total energy supply.  
Equation (2.25) of \cite{Virga15}, which Virga derives from a Rayleigh dissipation principle, can also be obtained from a different rearrangement of the energy jump condition \eqref{energy_jump}.
Substituting for $\tfrac{1}{2}\add{\partial_t\bX}$ from \eqref{spatial_velocity} into equation (\ref{energy_rewritten_2}), and using \eqref{energy_invariant_raw} and \eqref{energy_measure_final},
\begin{align}
&0=\tilde{E} - \bP\cdot\bv_0 + \partial_t s_0\tfrac{1}{2}\bP\cdot\add{\partial_s\bX} - \tfrac{1}{4}\partial_t s_0 \jump{\sigma}\norm{\jump{\partial_s\bX}}\, ,\label{virga_1}
\end{align}
then using \eqref{momentum_jump_basis} to insert for $\bP$ in the third term on the right, and the identity $\norm{\add{\partial_s\bX}}+\norm{\jump{\partial_s\bX}} = 4$,
\begin{align}
%\implies&\tilde{E} - \bP\cdot\bv_0 - \tfrac{1}{4}\partial_t s_0 \jump{\sigma} \norm{\add{\partial_s\bX}} - \tfrac{1}{4} \partial_t s_0 \jump{\sigma}\norm{\jump{\partial_s\bX}} \, ,\label{virga_2}\\
&0=\tilde{E} - \bP\cdot\bv_0 - \partial_t s_0 \jump{\sigma} \, .\label{virga_3}
\end{align}
By defining the dissipation $\mathscr{D} \equiv -\tilde{E} + \bP\cdot\bv_0$, one obtains Virga's equation (2.25) in \cite{Virga15}.
Equation (\ref{virga_3}) also corresponds to equation (34) of O'Reilly \cite{OReilly07} without director terms. The quantity $\mathsf{B}$ there is called the source of ``material momentum'', and in our example would correspond to $\jump{\sigma}$ in \eqref{virga_3}.  In another paper \cite{OReilly2007}, O'Reilly derives an additional configurational momentum balance law and shows that, given this additional balance, the energy jump condition reduces to an identity relating the power expended by the singular supplies of linear and material momentum.  Similarly, if one uses the momentum and energy balances, O'Reilly's configurational balance reduces to an identity.

%There is some experimental evidence for both link-type \cite{Gobat2001} and ball-type \cite{HammGeminard10} chains that dynamics and dissipation are not much affected by whether impact occurs with a hard or soft surface; it's not clear what, if anything, this implies.

Finally, we write general expressions for the tensions in the vicinity of the shock in terms of the singular supplies and the geometry.
Consider the momentum jump condition in the form \eqref{momentum_jump_basis}. Projecting it onto the jump in tangents and rearranging, we obtain the tension sum
\begin{align}
\add{\sigma}=2\mu\partial_t s_0^2-\frac{2\bP\cdot\jump{\partial_s\bX}}{\norm{\jump{\partial_s\bX}}}\, .\label{sum_of_tension}
\end{align}
The tension jump can be obtained from the energy jump in the form \eqref{energy_measure_final},
\begin{align}
\jump{\sigma}=\frac{4E}{\partial_t s_0 \norm{\jump{\partial_s\bX}}}\, .\label{jump_in_tension}
\end{align}
Thus, the tensions on either side of the shock are
\begin{align}
\sigma^\pm &= \mu\partial_t s_0^2 - \frac{\bP \cdot \jump{\partial_s\bX}}{\norm{\jump{\partial_s\bX}}} \pm \frac{2E}{\partial_t s_0 \norm{\jump{\partial_s\bX}}}\, . \label{sigma}
\end{align}
These expressions will be used in Appendix \ref{falling}.

\section{A compatibility relation between the supplies}\label{compatibility}

The momentum and energy supplies cannot be arbitrarily prescribed. 
Eliminating $\jump{\sigma}$ between \eqref{momentum_jump_basis} and \eqref{energy_measure_final}, we obtain the following relation between the supplies,
\begin{align}
E\norm{\add{\partial_s\bX}}=-\tfrac{1}{2}\partial_t s_0 \bP\cdot\add{\partial_s\bX}\norm{\jump{\partial_s\bX}}\, .\label{compatibility_relation}
\end{align}
This relation places no assumptions on the kinematics of the system, and is applicable to steady or unsteady motions.  We will use it in Section \ref{admissibility} to derive qualitative results about the admissibility of geometric discontinuities during pick-up and impact. 
Here we make a couple of general observations, valid as long as the body is not perfectly folded back, in which case $\add{\partial_s\bX}$ and both sides of the relation would all vanish.

Given a non-zero jump in tangents $\jump{\partial_s\bX}$, it is impossible to inject momentum into the shock without injecting or dissipating energy unless the momentum injection $\bP$ is orthogonal to the sum of tangents $\add{\partial_s\bX}$, that is, in the direction of the jump.  O'Reilly and Varadi \cite{OReillyVaradi04} have observed that this fact is inconsistent with some assumptions appearing in the literature on space-fixed loads acting on axially moving or rotating structures.

Given that the sign of the shock speed depends on a physically irrelevant parameterization, the sign of the energy supply $E$ depends only on the orientation of the momentum supply $\bP$ relative to the sum of tangents $\add{\partial_s\bX}$.
Letting $\partial_t s_0$ be negative, as it would be for an axially moving string parameterized in the direction of the axial flow, a positive projection $\bP\cdot\add{\partial_s\bX}$ would imply a supply of energy at the shock, a negative projection dissipation.

\section{Shocks conserving translation-invariant energy}\label{conserving}
Before considering dissipative shocks, it is worthwhile to look at cases where $E=0$. As mentioned earlier, equation \eqref{energy_measure_final} for the energy jump indicates that either the tension or the tangents must be continuous if $E=0$, that is,
\begin{align}
\jump{\sigma} &= 0 \, , \quad \mathrm{or}  \label{jump_sigma_zero}\\ 
\quad\jump{\partial_s\bX}&=0\, .\label{jump_tangent_zero}
\end{align}
Meanwhile, the compatibility relation \eqref{compatibility_relation} indicates that at least one of the following conditions must hold if $E=0$.  Either
\begin{align}
\bP & = 0\quad\text{if}\quad\add{\partial_s\bX}\ne 0\, , \quad \mathrm{or}\label{momentum_jump_zero}\\
\bP\cdot\add{\partial_s\bX} &= 0\quad\text{if}\quad\bP\ne 0\, , \quad \mathrm{or}\label{momentum_projection_zero}\\
\jump{\partial_s\bX} &= 0\, .\label{jump_tangent_zero_2}
\end{align}

Conditions \eqref{jump_tangent_zero} and \eqref{jump_tangent_zero_2} are the same.  It turns out that conditions \eqref{jump_sigma_zero} and \eqref{momentum_projection_zero} are also equivalent, in that if $\jump{\sigma}=0$, equation \eqref{momentum_jump_modified} reduces to
$ \bP=-(\sigma-\mu\partial_t s_0^2)\jump{\partial_s\bX}$,
so that \eqref{momentum_projection_zero} holds by orthogonality of the sum and jump of tangents in our inextensible string.
Although the conditions (\ref{jump_sigma_zero}) and (\ref{momentum_projection_zero}) are equivalent, the latter is more useful. The momentum injection $\bP$ is often provided by an external obstacle or tool, so it may be possible to infer something about its orientation from the physical attributes of this object. Information about the tension on either side of the shock is usually harder to determine.
There exists a third possibility, given by equation (\ref{momentum_jump_zero}).  Here there is no supply of momentum at the shock, and the sum of tangents across the shock is non-zero, meaning that the body is not perfectly folded back on itself.
To understand this case better, let us look at the momentum jump condition written in the sum-jump basis \eqref{momentum_jump_basis}.  Assume a kink, such that the jump in tangents is also non-zero.
In this case, the only solutions for which $\bP=0$ would require $\sigma^\pm = \mu \partial_t s_0^2$.  This condition obtains for the lariats \cite{Routh55, Healey1990}, solutions that have fixed or rigidly translating shape and constant axial flow.  Kinks are compatible with these solutions \cite{OReillyVaradi04,Hanna15}.
%$\partial_t \bX = T \partial_s\bX$. Here $T$ is constant 

It must be emphasized that the condition $\bP = 0$ is not sufficient for conservation of translation-invariant energy. It can be inferred from \eqref{compatibility_relation} that a non-zero $E$ can still exist if the sum $\add{\partial_s\bX}=0$. The falling folded chain discussed in Appendix \ref{falling} is an example of such a case \cite{Heywood55,CalkinMarch89,Schagerl97,Virga15}.

 \section{Admissibility of a kink during pick-up and impact}\label{admissibility}
 
In this section, we use the compatibility relation \eqref{compatibility_relation} to determine whether a discontinuity in tangents is admissible in certain physical circumstances.
It is reasonable to assume that such a propagating kink will be dissipative, so that $E$ is non-positive.  Indeed, this assumption follows from a form of the Clausius-Duhem inequality for isentropic, isothermal, purely mechanical systems, as discussed in Appendix \ref{clausius_duhem}.  A more general system involving adhesion energy and a jump in entropy, which violates this assumption, is discussed in Appendix \ref{peel}.
For $E \le 0$ and a parameterization such that $\partial_t s_0<0$, we find 
\begin{align}
\bP\cdot\add{\partial_s\bX}\le 0 \, . \label{compatibility_inequality} 
\end{align} 
%with the equality holding for $E=0$.  
This inequality restricts the directions of $\bP$ relative to the sum of tangents. Since the momentum injection comes from the interaction of the string with an external source, the physical settings of the system may dictate whether this source is capable of providing a $\bP$ in a direction consistent with  \eqref{compatibility_inequality}.  For example, a smooth, frictionless, non-adhesive surface can only provide a $\bP$ pointing outwards in the direction of the surface normal.

The assumption of the non-positivity of $E$, as defined in \eqref{energy_invariant_definition}, can be examined in the light of the total energy balance \eqref{energy_rewritten_2}.  The implication is that $\tilde{E}\le \tfrac{1}{2}\bP\cdot\add{\partial_t\bX}$, meaning that the rate of change of the total energy across the shock $\tilde{E}$ cannot be greater than the rate at which work is done on the material at the shock $\tfrac{1}{2}\bP\cdot\add{\partial_t\bX}$ by external forces. The difference between the two quantities tells us about the translation-invariant energy being dissipated in the shock.  %When the rate of work by external forces exactly balances the rate of change of the total energy, the shock does not dissipate energy.
 Note that given \eqref{momentum_jump_basis}, \eqref{compatibility_inequality} implies $\jump{\sigma} \ge 0$.  In an inextensible string, the tension cannot decrease across the shock; the tension at a material point passing through the shock cannot decrease.

\begin{figure}[h!]
	\begin{subfigure}[t]{8.5cm}
		\includegraphics[width=\textwidth]{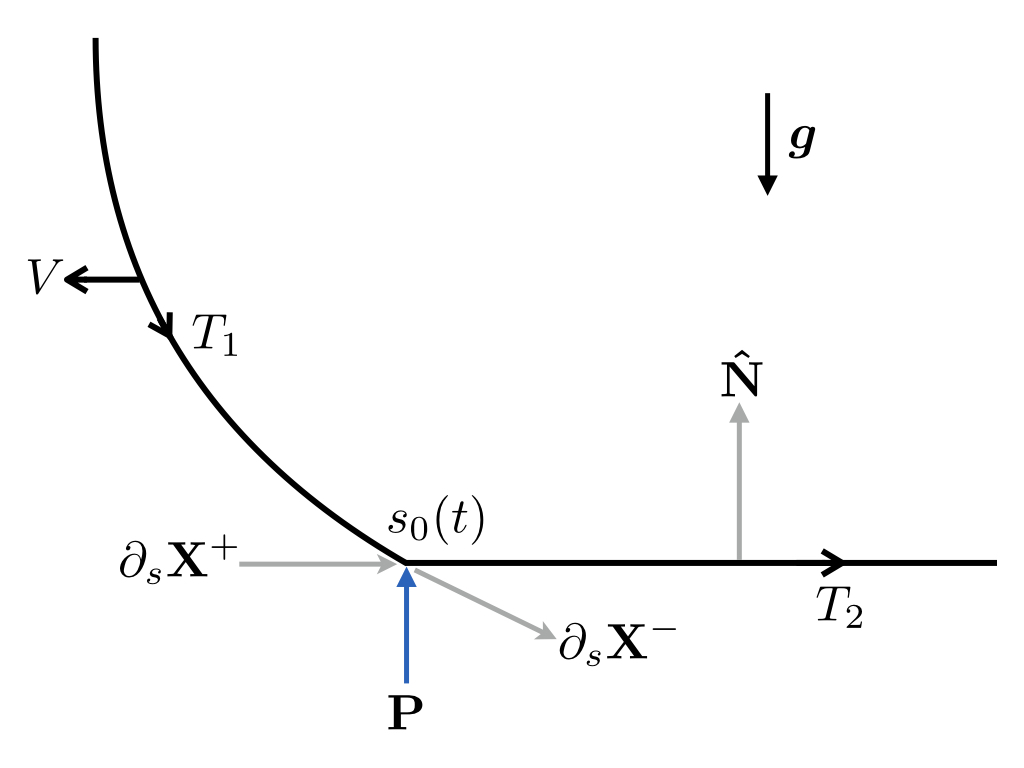}
		\captionsetup{justification=centering}
		\vspace{-.5cm}
		\caption{}
		\label{impact_a}
	\end{subfigure}
	\begin{subfigure}[t]{8.5cm}
		\includegraphics[width=\textwidth]{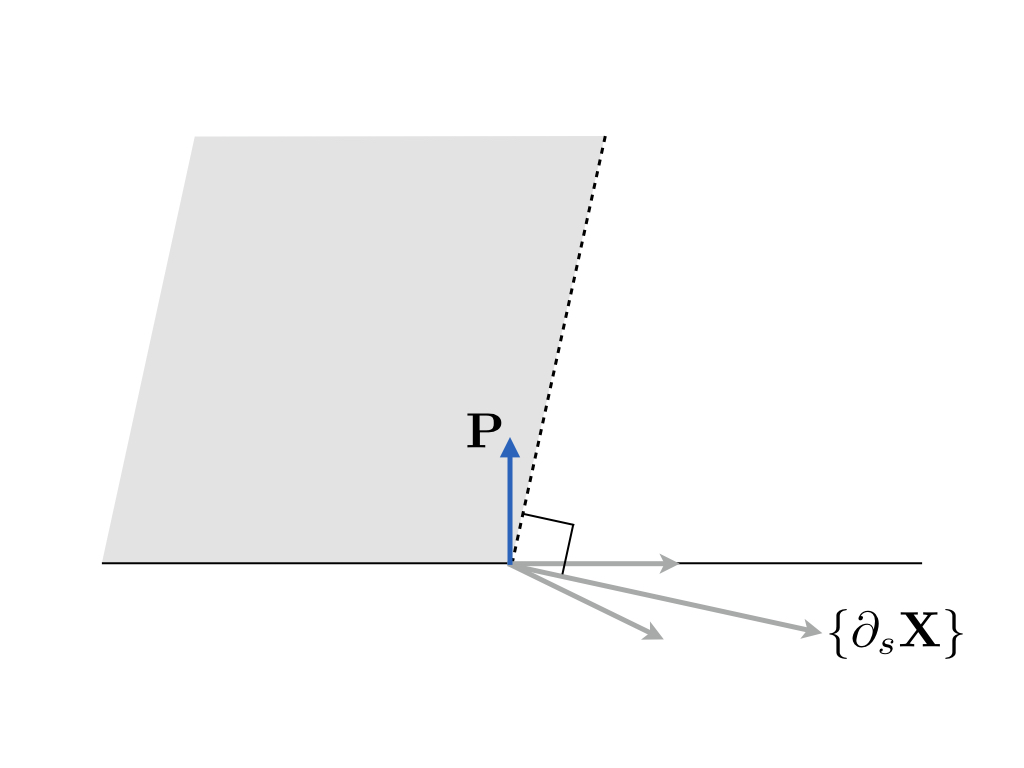}
		\captionsetup{justification=centering}
		\vspace{-.5cm}
		\caption{}
		\label{impact_b}
	\end{subfigure}
		\captionsetup{margin=.75cm}
	\caption{An axially moving string impacting a surface with unit normal $\uvc{N}$ antiparallel to gravity $\bg$. (\subref{impact_a}) The free piece translates with velocity $-V\partial_s\bX^+$ and flows with axial velocity $T_1\partial_s\bX^-$, while the contacting piece flows with axial velocity $T_2\partial_s\bX^+$.  A singular supply $\bP$ acts at the discontinuity at $s_0(t)$, which moves to the left as the free piece is laid down onto the surface.  (\subref{impact_b}) The sum of tangents and the grey region of admissibility where $\bP\cdot\add{\partial_s\bX} \le 0$ are shown.}
	\label{impact}
\end{figure}

\begin{figure}[h!]
	\begin{subfigure}[t]{8.5cm}
		\includegraphics[width=\textwidth]{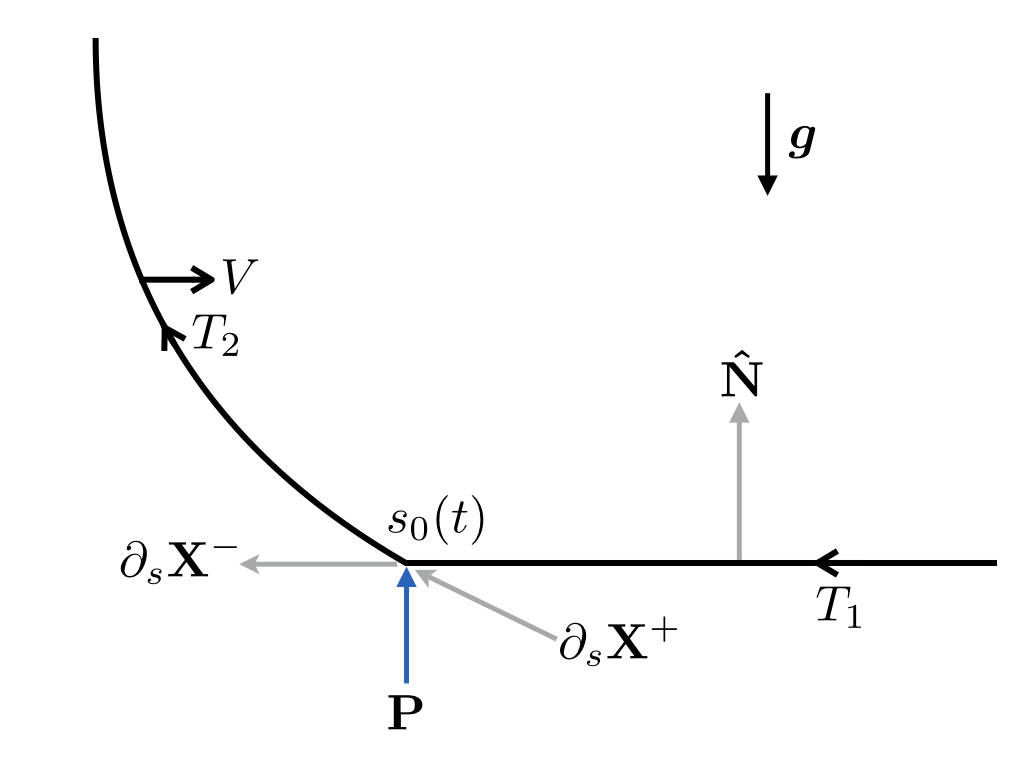}
		\captionsetup{justification=centering}
		\vspace{-.5cm}
		\caption{}
		\label{pick-up_a}
	\end{subfigure}
	\begin{subfigure}[t]{8.5cm}
		\includegraphics[width=\textwidth]{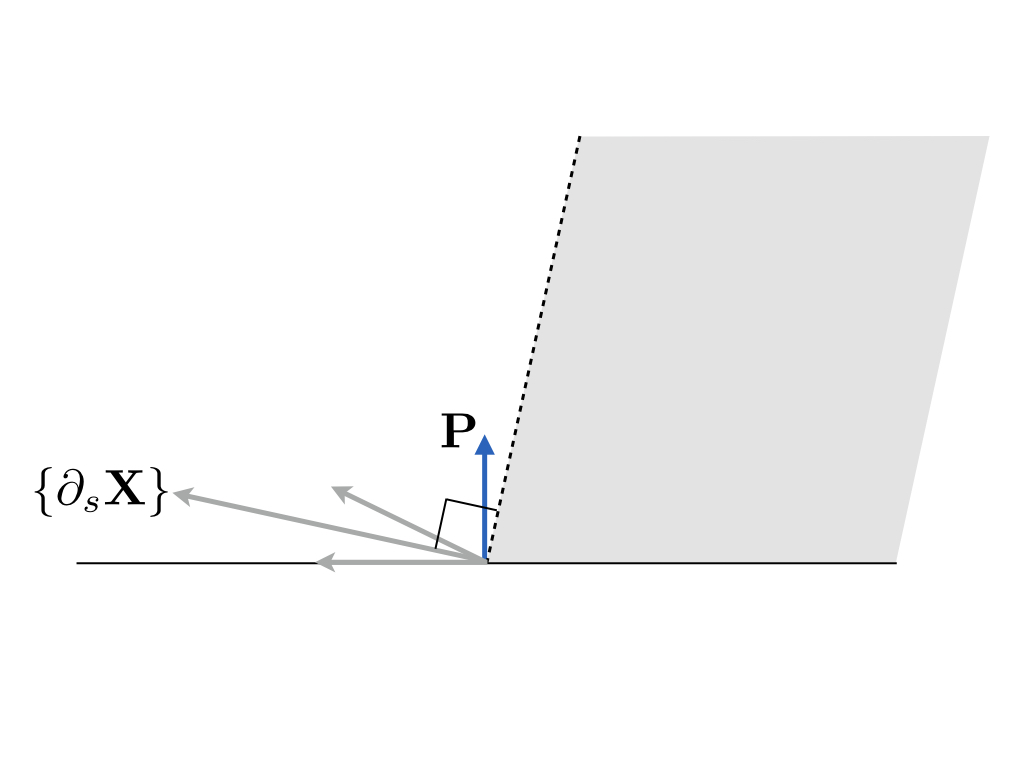}
		\captionsetup{justification=centering}
		\vspace{-.5cm}
		\caption{}
		\label{pick-up_b}
	\end{subfigure}
	\captionsetup{margin=.75cm}
	\caption{An axially moving string picked up from a surface with unit normal $\uvc{N}$ antiparallel to gravity $\bg$. (\subref{impact_a}) The free piece translates with velocity $-V\partial_s\bX^-$ and flows with axial velocity $T_2\partial_s\bX^+$, while the contacting piece flows with axial velocity $T_1\partial_s\bX^-$.  A singular supply $\bP$ acts at the discontinuity at $s_0(t)$, which moves to the right.  (\subref{impact_b}) The sum of tangents and the grey region of admissibility where $\bP\cdot\add{\partial_s\bX} \le 0$ are shown.}
	\label{pick-up}
\end{figure}

We now apply condition \eqref{compatibility_inequality} to determine the admissibility of kinks at the pick-up and impact points of a string interacting with a smooth, rigid plane normal to gravity.  We include gravity simply to preclude a lariat-type solution in which the reaction stress $\bP$ vanishes.  Figures \ref{impact} and \ref{pick-up} show the kinematics of these closely related problems.  Without loss of generality, we consider an axially flowing string with an additional rigid translation imposed on the free piece, and parameterize the string in the direction of the axial flow for each case.
  Defining the axial speeds $T_1$ and $T_2$ and the translational speed $V$ as non-negative, the spatial velocity of the non-material kink is to the left in Figure \ref{impact_a} (impact) and to the right in Figure \ref{pick-up_a} (pick-up).  
The material velocities in the vicinity of the impact point are $\partial_t\bX^- = T_1\partial_s\bX^- - V\partial_s\bX^+$ and $\partial_t\bX^+ = T_2 \partial_s\bX^+$.  Equation \eqref{velocity_constraint} requires $T_1 = T_2 + V$, and equation \eqref{shock_speed} then gives the shock velocity $\partial_t s_0 = - T_1$ in the material coordinates.
The material velocities in the vicinity of the pick-up point are $\partial_t\bX^- = T_1\partial_s\bX^- $ and $\partial_t\bX^+ = T_2 \partial_s\bX^+ - V\partial_s\bX^-$.   Equation \eqref{velocity_constraint} requires $T_2 = T_1 + V$, and equation \eqref{shock_speed} then gives the shock velocity $\partial_t s_0 = - T_2$ in the material coordinates.

The grey regions in Figures \ref{impact_b} and \ref{pick-up_b} are regions of admissibility, where $\bP\cdot\add{\partial_s\bX}$ is non-positive.  The plane can provide only a normal reaction force $\bP$.  The tangents lie in the half-space above the plane, and for impact (Figure \ref{impact}) it is clear that $\bP$ is always admissible, and energy will be dissipated.  However, during pick-up (Figure \ref{pick-up}), the projection $\bP\cdot\add{\partial_s\bX}$ is always non-negative.  A positive projection would imply an unphysical injection of energy at the kink.  There are two ways to resolve this difficulty.  One is if the surface has a bump at the pick-up point that could provide a $\bP$ that points into the grey region.  The other is if the reaction stress $\bP$ vanishes.  Excluding the perfectly folded case, this and the relation \eqref{compatibility_relation} would imply that $E=0$.  From the discussion in Section \ref{conserving}, we know that this implies that either the tension or tangents are continuous, $\jump{\sigma}=0$ or $\jump{\partial_s\bX}=0$.  Here we consider both possibilities, and show that in the presence of gravity and a vanishing $\bP$, each implies the other in steady-state.
Assuming first the continuity of tension, and no momentum injection, the momentum balance \eqref{momentum_jump_basis} during pick-up becomes $-(\sigma - \mu T_2^2)\jump{\partial_s\bX}=0$.  In the presence of gravity, the steady-state bulk tension in an axially flowing string follows the catenary solution $\sigma = \mu T_2^2 - \left({\bf g}\cdot\uvc{n}/\kappa\right)$ where both the curvature $\kappa$ and the projection of gravity $\bg$ onto the curve normal $\uvc{n}$ are always non-zero.  Hence $\sigma - \mu T_2^2 \ne 0$, so it must follow that $\jump{\partial_s\bX} = 0$.  For an unsteady instantaneous configuration with a kink, gravity will be balanced by an acceleration that will smooth the kink.  On the other hand, if we assume continuity of tangents, the momentum balance \eqref{momentum_jump_basis} becomes $-\jump{\sigma}\partial_s\bX=0$, and thus $\jump{\sigma}=0$ for both steady and unsteady configurations.  Note that we can say more.  If the tangents are continuous, the momentum balance \eqref{momentum_jump_basis} says that any nonzero $\bP$ must be parallel to the sum of tangents, an impossibility if $\bP$ is normal to the surface.  Hence, for any steady or unsteady smooth pick-up or impact from a smooth, frictionless surface, both $\bP$ and $\jump{\sigma}$ vanish.  From all of this, we see that the ``chain fountain'' and related problems are not so simple as one might hope (see Appendix \ref{heap}).

To summarize, a shock in the tangents of a string can only exist if the momentum supply satisfies the condition \eqref{compatibility_inequality}. For situations where a nonzero injection of momentum is required but cannot satisfy \eqref{compatibility_inequality}, including the pick-up of a string off of a smooth, frictionless surface, the tangents and the tension must be continuous at the contact discontinuity.  For a kink to exist at the pick-up point, a properly oriented obstacle is necessary.  An experimental illustration of this point will be shown shortly.

We have established a criterion for the admissibility of a kink.  We now discuss another criterion that predicts the existence of a kink, associated with non-vanishing $\bP$ and negative $E$.
Consider again the impact problem shown in Figure \ref{impact}. We know that the force $\bP$, if it exists, can only point in the direction normal to the surface. Projecting the momentum balance \eqref{momentum_jump_modified} onto the unit surface normal $\uvc{N}$, we obtain
\begin{align}
\bP\cdot\uvc{N} = -(\sigma^- - \mu \partial_t s_0^2)\sin\theta\, ,\label{projection_momentum}
\end{align}
where $\cos\theta = \partial_s\bX^+\cdot\partial_s\bX^-$.  For a non-adhesive surface, $\bP\cdot\uvc{N}$ must be non-negative.  A kink will have some angle $0 < \theta < \pi$, excluding the perfectly folded case.  If $\sigma^- - \mu \partial_t s_0^2\le 0$, $\bP\cdot\uvc{N}\ge 0$ and a kink is admissible.  If $\sigma^- - \mu \partial_t s_0^2 >0$, a kink is inadmissible, $\theta = 0$ and $\bP\cdot\uvc{N}=0$.  The folded case $\theta=\pi$ is an exception, but not relevant to the impact geometry.  %; we discuss a falling folded chain in Appendix \ref{falling}.
An impact force and associated kink will exist if $|\partial_t s_0| > \sqrt{\sigma^- / \mu}\,$, that is, if the material velocity of the non-material impact point is greater than the local transverse wave speed in the string on the pre-impact side.  Because $\partial_ts_0 = -T_2$ for impact, and $\sigma < \mu T_2^2$ for a concave-down catenary arch, this condition always holds for such a steady state.  However, it need not hold for transient configurations.
Burridge and co-workers \cite{Burridge1982} derived a similar criterion for the existence of a reaction force at the wrapping impact point for the geometrically linearized problem of a stretched string vibrating against a curved obstacle, as happens in certain Indian musical instruments.  They simply state without justification that this criterion cannot be met for the corresponding unwrapping motion. %IN CONTRAST WITH IMPULSIVE LOADING LITERATURE? 

\begin{figure}[t!]
	
	\begin{subfigure}[t]{\textwidth}
		\includegraphics[width=15cm]{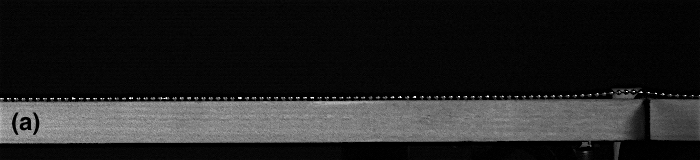}
		\vspace{-10pt}
	\end{subfigure}
		
	\begin{subfigure}[t]{\textwidth}
		\includegraphics[width=15cm]{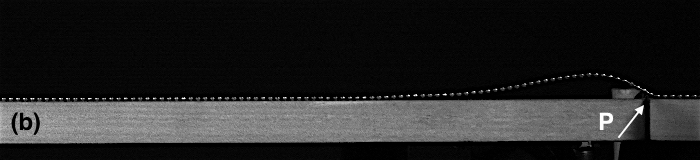}
		\vspace{-10pt}
	\end{subfigure}	
	
	\begin{subfigure}[t]{\textwidth}
		\includegraphics[width=15cm]{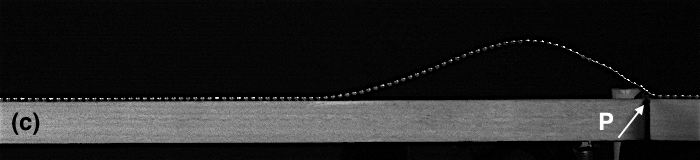}
		\vspace{-10pt}
	\end{subfigure}
	
	\begin{subfigure}[t]{\textwidth}
		\includegraphics[width=15cm]{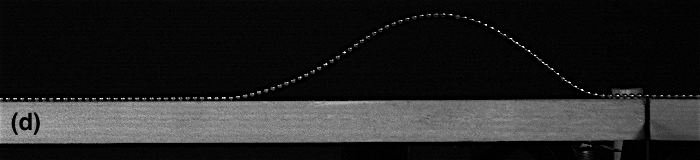}
		\vspace{-10pt}
	\end{subfigure}
	
	\begin{subfigure}[t]{\textwidth}
		\includegraphics[width=15cm]{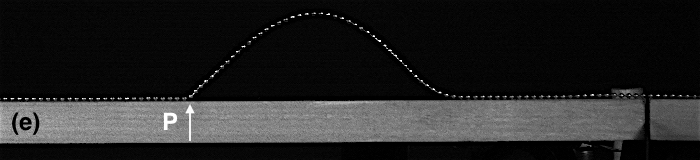}
		\vspace{-10pt}
	\end{subfigure}
	
	\begin{subfigure}[t]{\textwidth}
		\includegraphics[width=15cm]{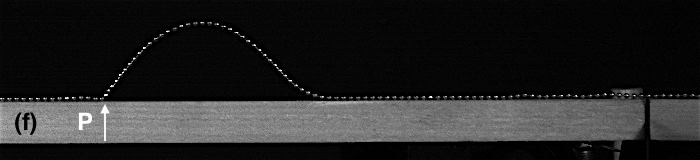}
		\vspace{-10pt}
	\end{subfigure}
	\captionsetup{margin=1.5cm}
	\vspace{-.2cm}
	\caption{Still images from a video of pick-up and impact discontinuities.  A ball-and-link chain is pulled at several m/s to to the left over a smooth surface.  In frames (a)-(c), a small blade protrudes above the surface on the right, causing a ``bubble'' to form.  After the blade is retracted, the shape propagates freely.  Kinks can be seen at the pick-up points in frames (b)-(c) and at impact in frames (e)-(f).  Inferred reaction forces are labeled.}
	\label{bubble}
\end{figure}

Gobat and Grosenbaugh's study of the touchdown region of catenary moorings  \cite{Gobat2001} featured experiments on submerged link chains vertically vibrated against a platform.  They present data on the shape of the chain during this process which clearly show the disappearance of a kink during the transition to upward motion of the chain, and smooth pick-up points during upward motion.  Kinks appear during sufficiently rapid downward motion.  
Additional qualitative support for our predictions comes from the video stills shown in Figure \ref{bubble}.  This laboratory demonstration of a pair of propagating contact discontinuities is essentially the same setup as in \cite{Hanna14}.  A ball-and-link chain is pulled at several m/s to the left over a smooth surface.  A retractable blade near the right hand side creates a ``bubble'' in the chain.  After the blade retracts, the disturbance moves freely to the left and eventually dies down.  While this is not a steady-state configuration, the tangential speed of the chain is nearly constant, and is quite rapid compared to the shape dynamics.  Many of the features we discuss above are apparent.  The blade provides a sideways reaction force that can sustain a pick-up kink; immediately once it is retracted, the pick-up point becomes smooth.  Later, the impact point undergoes a transition from smooth to kinked.

\section{Further discussion and conclusions}\label{discussion}

We have examined a simple mechanical system, and made a surprising prediction that appears to have gone unnoticed in the literature, but is consonant with previous experimental observations.  
A kink in a flexible structure is generally inadmissible during pick-up.
This statement follows from a compatibility relation  \eqref{compatibility_relation} between stress and translation-invariant power supplies, along with basic assumptions about the reaction stress, at the contact discontinuity.  While our conclusion is based on excluding exceptional solutions for which there is no reaction stress (including lariat-type solutions), which we have only been able to do explicitly for steady-state configurations, the limited experimental evidence suggests that the conclusions are more broadly applicable.

At the discontinuity, we supplemented momentum conservation with energy conservation, while allowing for source terms.  These balances correspond to symmetries in spatial and temporal coordinates.  An alternative is to use a balance of some configurational-like or ``material'' quantity \cite{Pede_peeling2006, OReilly07, OReilly2007}, corresponding to a boundary shift or a symmetry in material coordinates.

The simple example of an inextensible string allowed us to use a very convenient orthogonal pair, the sum and jump of tangents, to transparently express vector quantities and perform calculations.
A non-orthogonal pair, the sum and jump of velocities, was used to isolate the translation-invariant part of the energy.
A natural consequence of our energy splitting is that the external stress supply is paired with the average velocity across the kink, in keeping with standard treatments of particle impact mechanics.  This leads to the compatibility relation used in our analysis.

An immediate question is, how do these concepts extend to more complicated one-dimensional bodies?  We expect that axial stretching will complicate the problem without providing much new physics, at least for a hyperelastic material with convex elastic energy.  However, higher-order discontinuities in rods and other structures with bend and twist resistance are potentially interesting.  Recent experiments on ``torsional locomotion'' show how such structures can be propelled away from partial constraints by jumps in curvature or twist \cite{Bigoni14}.  
Rotational invariance properties of the power dissipation terms will likely be important for such systems.

Other ways of looking at these problems might consider the dynamics of the geometry, or of the non-material point of contact.  An example problem in which the latter type of formulation is possible, given certain assumptions about the geometry and dissipation, is discussed in Appendix \ref{falling}.

\appendix

\section{The Clausius-Duhem inequality}\label{clausius_duhem}

The conclusions made in this work about the admissibility of a shock in a string rely on the assumed non-positivity of the invariant measure of energy dissipation $E$ as defined in \eqref{energy_invariant_definition}. In this appendix, we briefly note that this assumption can be inferred from the second law of thermodynamics\footnote{We thank A. Gupta for suggesting this line of argument.}.
%The Clausius-Duhem inequality is a statement that the rate of change of the free energy of a system cannot exceed the power expended on it. 
For our simple string, we may write
\begin{align}
%\frac{d}{dt}\int_{s_1}^{s_2} ds\, \left(\Psi+\tfrac{1}{2}\mu\partial_t\bX\cdot\partial_t\bX\right) \le \sigma\partial_s\bX\cdot\partial_t\bX\mid_{s_1}^{s_2} + \tfrac{1}{2}\bP\cdot\add{\partial_t\bX}\, ,\label{clausius_duhem_general}
\frac{d}{dt}\int_{s_1}^{s_2} ds\, \tfrac{1}{2}\mu\partial_t\bX\cdot\partial_t\bX \le \left. \sigma\partial_s\bX\cdot\partial_t\bX \,\right|_{s_1}^{s_2} + \tfrac{1}{2}\bP\cdot\add{\partial_t\bX}\, ,\label{clausius_duhem_general}
\end{align}
%where $\Psi$ is the free energy of the system. For the present case of purely mechanical inextensible strings we set $\Psi=0$. 
%The two terms on the right side of the inequality are the power expended by the contact force and the point force $\bP$ on the material. 
for a material interval enclosing a moving discontinuity at $s_0(t)$, such that $s_1< s_0(t) <s_2$.  This form of the Clausius-Duhem inequality, without the final term involving the stress source $\bP$, can be arrived at from an energy balance and entropy imbalance following Chapters 6 \& 7 of Gurtin \cite{Gurtin00} and noting that our string has no free energy--- neither internal energy nor entropy terms exist.
  We split the integral over two time-dependent intervals, apply the Leibniz rule, and let $s_1$ and $s_2$ approach $s_0$ from below and above, respectively, to obtain
\begin{align}
-\jump{\sigma\partial_s\bX\cdot\partial_t\bX + \tfrac{1}{2}\mu \partial_t s_0 \partial_t\bX\cdot\partial_t\bX} - \tfrac{1}{2}\bP\cdot\add{\partial_t\bX} &\le 0\, ,\label{clausius_duhem_1}
\end{align}
which from \eqref{energy_jump} and \eqref{energy_invariant_definition} is simply
\begin{align}
%\tilde{E} - \tfrac{1}{2}\bP\cdot\add{\partial_t\bX} &\le 0\, .\label{clausius_duhem_final}
E &\le 0\, .\label{clausius_duhem_final}
\end{align}
For a rod, an internal bending energy term and the working of an areal torque source need to be considered as well.

\section{Falling folded chain}\label{falling}

\begin{figure}[h!]
	\includegraphics[width=4cm]{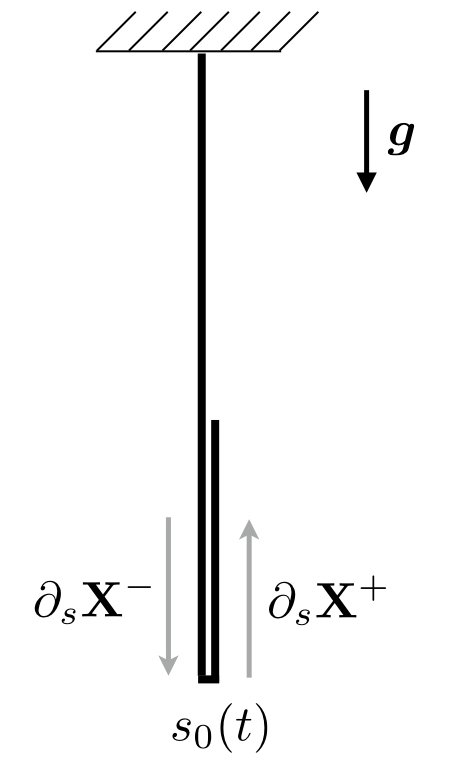}
	\captionsetup{margin=6cm}
	\caption{A folded chain with one fixed and one falling segment.}
	\label{falling_folded_chain}
\end{figure}

Consider a chain fixed at one end $s=0$, while the other end $s=l$ falls after being released from rest in a perfectly folded configuration (Figure \ref{falling_folded_chain}).  
This problem has a history as a challenging and counterintuitive example in dynamics \cite{Satterly51, Heywood55, CalkinMarch89, Schagerl97, Tomaszewski06, WongYasui06, Virga15}, and also sees application in the context of cable deployment from orbiting spacecraft \cite{KaneLevinson77} and recreational bungee jumping (\cite{Heck10} and references therein).  From the perspective of the current paper, the perfectly folded chain is a special configuration where the sum of tangents $\add{\partial_s\bX} = 0$.  Additionally, there is no external source $\bP$, and our compatibility relation \eqref{compatibility_relation} is satisfied identically for any arbitrary choice of internal dissipation $E = \tilde{E}$.

We will combine the general expressions \eqref{sigma} with a constitutive prescription for $E$ to recover Virga's \cite{Virga15} results for this problem.  The prescription may be derived from O'Reilly and Varadi's general prescription for the jump in tension in equation (7.7) of \cite{OReillyVaradi99}, %\footnote{ in light of later work \cite{OReilly07}the prescription is for the point source of a generalized quantity called the ``material momentum", which in our case of an inextensible string reduces to the jump in the tension $\jump{\sigma}$.}.
\begin{align}
\jump{\sigma}=-2 f \mu |\partial_t s_0| \partial_t s_0 \, ,\label{jump_prescription}
\end{align}
where $f$ is a constitutive parameter. Substituting \eqref{jump_prescription} in \eqref{energy_measure_final} prescribes $E$ as
\begin{align}
E=-\tfrac{1}{2} f \mu  |\partial_t s_0| (\partial_t s_0)^2 \norm{\jump{\partial_s\bX}}\, ,\label{energy_prescription_1}
\end{align}
or, using velocity compatibility \eqref{velocity_compatibility}, as
\begin{align}
E=-\tfrac{1}{2} f \mu |\partial_t s_0| \norm{\jump{\partial_t \bX}}\, ,\label{energy_prescription}
\end{align}
which agrees with equation (9) of \cite{Virga14} and equation (4.1) of \cite{Virga15}.

With $\bP$ set to zero and a constitutive prescription of $E$ specified, the bulk equation $\mu\partial^2_t\bX=\partial_s(\sigma\partial_s\bX)+\mu\bg$ can be projected onto the tangents on each side and integrated, with \eqref{sigma} providing the boundary conditions at the fold.
Using \eqref{energy_prescription_1}, we obtain the boundary conditions
\begin{align}
\sigma^\pm &= \mu \partial_t s_0\left( \partial_t s_0 \mp f |\partial_t s_0| \right)\, .\label{falling_chain_sigma}
\end{align}
For these kinematics, velocity compatibility \eqref{velocity_compatibility} indicates that on the falling side, $\partial_t\bX=-2\partial_t s_0\partial_s\bX$, and thus $\partial_t^2\bX=-2\partial_t^2s_0\partial_s\bX$.  Integrating the projected bulk equation $-2\mu\partial_t^2s_0 = \partial_s\sigma - \mu | \bg |$ and using $\sigma(l)=0$ at the free end, we obtain
\begin{align}
\sigma(s)=-\mu( | \bg | -2\partial_t^2 s_0)(l-s)\quad\text{for}\quad s_0\le s\le l\, .\label{bulk_solution_falling_part}
\end{align}
Evaluating this at $s=s_0$ and using \eqref{falling_chain_sigma}, we may write a differential equation for the dynamics of the shock,
\begin{align}
\partial_t s_0(\partial_t s_0 - f |\partial_t s_0|) + ( | \bg | - 2 \partial_t^2 s_0) (l-s_0) = 0 \, ,\label{shock_governing_equation}
\end{align}
in agreement with the free-end case of equation (6.6) of \cite{Virga15}.  Similarly, the % the dynamic weight of the chain, or the 
reaction force at the support can be obtained by integrating the bulk equation on the stationary side and employing the boundary condition at the fold.  We obtain
\begin{align}
\sigma(0) = \mu \partial_t s_0 (\partial_t s_0 + f |\partial_t s_0|) + \mu | \bg | s_0 \, ,\label{falling_chain_dynamic_weight}
\end{align}
in agreement with equation (6.7) of \cite{Virga15}.

The freedom to prescribe $E$ is a special feature of the perfectly folded chain that cannot be used in more realistic configurations such as those of the experiments in \cite{Tomaszewski06}.

\section{Peeling}\label{peel}

\begin{figure}[h!]
\centering
		\includegraphics[width=8cm]{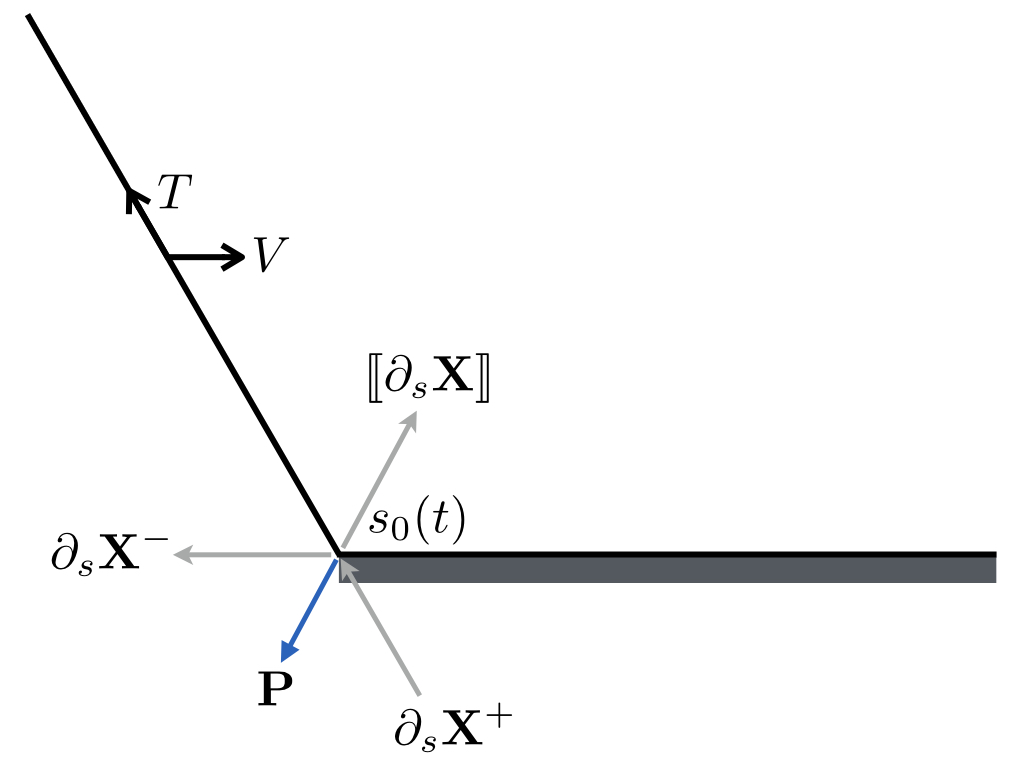}
		\captionsetup{margin=4cm}
	\caption{A peeling tape, akin to the pick-up problem of Figure \ref{pick-up} without gravity and with a stationary contacting piece adhered to the surface.  The reaction stress $\bP$ points antiparallel to the jump in tangents.}\label{Peeling_full}
\end{figure}

Peeling \cite{Rivlin44, Kendall75, Ericksen1998, BurridgeKeller1978} is a moving-boundary problem related to standard adhesion tests, and is sometimes used as a toy model of fracture.
Interesting inertial and material instabilities have been observed in peeling tapes \cite{Cortet13, Dalbe14}.  In this appendix, we briefly show how classical quasi-static results relate to our approach.
 
Consider a tape peeled from a plane rigid surface, as in Figure \ref{Peeling_full}.  The stress supply $\bP$ for an adhesive surface can point in any direction.  
To obtain the power balance, we modify the Lagrangian density \eqref{lagrangian_density} to be 
\begin{align}
2\mathcal{L}=\mu\partial_t\bX\cdot\partial_t\bX - \sigma (\partial_s \bX \cdot \partial_s \bX-1) -  2\varepsilon\, , \label{peeling_lagrangian_density}
\end{align}
where $\varepsilon$ is the internal energy density of the system, a quantity we will relate to the adhesion energy below.  We don't concern ourselves with constraint terms on the position of the adhered portion of the tape, as they would not end up in the jump condition.
 The jump condition \eqref{energy_jump} is modified accordingly,
\begin{align}
\tilde{E} + \jump{\sigma \partial_s\bX \cdot \partial_t \bX + \partial_t s_0 \left( \tfrac{1}{2} \mu \partial_t \bX \cdot \partial_t \bX + \varepsilon \right)} = 0\, .\label{peeling_energy_jump}
\end{align}
%equation (21) of \cite{Pede_peeling2006}, where the ``tip force $S$" corresponds to $\jump{\varepsilon}$ in (\ref{peeling_energy_jump}) above. 
Note that in a typical analysis of this problem, $\partial_t\bX^-=0$.  From the pick-up kinematics (Section \ref{admissibility}), we know then that $\partial_t\bX^+=-\partial_ts_0\jump{\partial_s\bX}$.
Let us write
\begin{align}
\jump{\varepsilon} = 0 - (-\Gamma /d) = \Gamma /d\, ,\label{jump_in_adhesion_energy}
\end{align}
where $\Gamma$, a positive number, is the surface adhesion energy and $d$ is the thickness of the tape.
 In the quasi-static limit ($\mu=0$), the assumption that $\tilde{E}=0$ along with the association 
$\sigma^+=F/bd$, where $F$ is the magnitude of the axial pull force and $b$ is the width of the tape, recovers the inextensible limit of the Rivlin-Kendall result \cite{Rivlin44, Kendall75}
\begin{align}
\Gamma =  (F/b) (1 - \cos\theta)\, ,\label{kendall_comparison_equation}
\end{align}
where $\cos\theta = \partial_s\bX^+\cdot\partial_s\bX^-$.  Note two special cases.  If the peeling angle $\theta=0$, peeling cannot occur, while if the adhesion energy $\Gamma = 0$, the peeling angle must be zero, a situation akin to our pick-up problem in Section \ref{admissibility}.

If we rearrange  \eqref{peeling_energy_jump} in the manner performed in the main text, we obtain
\begin{align}
E +  \partial_t s_0 \jump{\varepsilon} = \tfrac{1}{4} \partial_t s_0 \jump{\sigma} \norm{\jump{\partial_s\bX}}\, .\label{adhesion_energy_balance}
\end{align}
We now recognize that $E + \partial_t s_0 \jump{\varepsilon} = 0$, meaning that the peeled tape has a higher translation-invariant energy density than the adhered tape, corresponding to an ``injection'' of energy $E>0$.  Thus, the tension $\sigma$ is continuous for a kink, and  from \eqref{momentum_jump_basis} we see that $\bP$ points in the direction of the jump in tangents (antiparallel with the jump for a positive tension, as shown in Figure \ref{Peeling_full}).
The compatibility relation of Section \ref{compatibility} will be modified to 
\begin{align}
\big(E + \partial_t s_0\jump{\varepsilon}\big) \norm{\add{\partial_s\bX}} = - \tfrac{1}{2} \partial_t s_0 \bP \cdot \add{ \partial_s \bX } \norm{\jump{\partial_s\bX}}\, ,\label{peeling_compatibility_relation}
\end{align}
which is satisfied by having $E + \partial_t s_0 \jump{\varepsilon} = 0$ and $\bP \cdot \add{ \partial_s \bX } = 0$.
  
The quasi-static Rivlin-Kendall result was also obtained by Pede and co-workers \cite{Pede_peeling2006}, who employed a configurational balance instead of an energy balance.  Our energy approach, involving consideration of the invariant quantity $E$, is consistent with the main result and also gives us the direction of $\bP$, but we only obtain the main result (as well as the magnitude of $\bP$) by considering $\tilde{E}$.  This is perhaps to be expected, as the adhesive surface breaks the translational symmetry of the problem.  Application of an isothermal Clausius-Duhem inequality to the shock would tell us that our positive $E$ needs to be compensated by a jump in entropy.

\section{String pulled from a pile}\label{heap}

Variations on another classic variable-mass problem \cite{OReillyVaradi99} that have attracted much recent interest involve the deployment of a string-like body from a loose, stationary pile.  
As these variations are often improperly conflated, we briefly summarize known results here.  

Two types of chain are typically used in the relevant experiments.  A link chain offers essentially no bending resistance, though may offer some hard-wall twist resistance if sufficiently twisted.  A ball-and-link chain offers a hard-wall bending resistance if sufficiently bent.  
Pulling one end of either type of chain with constant velocity leads to the formation of a small arch-like structure near the pile.  The ball-and-link chain pushes back on the pile \cite{GeminardPalacio}, while the link chain does not \cite{HannaKing11}.   If the ball-and-link chain's deployment is an accelerated motion driven by gravity, a large arch known as a ``chain fountain'' forms \cite{MouldSiphon2, Biggins14}.  If observed for long enough, the structure resembles a very noisy catenary.  The effect is small, if present at all, in a link chain, and so appears to require the existence of a push-back effect arising from bending resistance \cite{Biggins14}.  
The smaller arch structure formed by either type of chain has been observed on short time scales, where gravity may be ignored.  Though it does not require a bending push-back, we speculate that its formation may be connected to the twist resistance of the chain.
We recall from Section \ref{admissibility} that the pick-up and impact effects discussed in the present paper have been observed in both types of chain.

Both Biggins \cite{Biggins14} and Virga \cite{Virga14} have proposed models of the chain fountain which are not consistent with the analysis in the present paper.  They make assumptions to simplify the geometry and kinematics of the system which violate velocity compatibility (and thus indirectly conservation of mass) and/or require energy injection.  Both authors seem aware of these limitations, and our present treatment is also unable to capture the experimental observations of these deployment phenomena.  Likewise, it was recognized by O'Reilly and Varadi \cite{OReillyVaradi99} that textbook treatments of similar string problems lead to impossibilities when they beg the question of the geometry and kinematics.  It is clear from the analysis in this paper that the question of matching a concave-down catenary with a surface-contacting piece is not trivial.  We conclude that the geometry of these problems is likely to be a dynamic, three-dimensional one, as already suggested by experiments \cite{HannaKing11, Martins16}.

\section*{Acknowledgments}
The chain bubble demo of Figure \ref{bubble} was designed and filmed by R. B. Warner during his time as an undergraduate researcher in our group.
We thank J. S. Biggins, E. Fried, A. Gupta, O. M. O'Reilly, and E. G. Virga for helpful and detailed discussions.  This work was supported by U.S. National Science Foundation grant CMMI-1462501.

\bibliographystyle{unsrt}
%\bibliography{refs_noncon}

\end{document}